# Mathematical models for pain: a systematic review

Victoria Ashley Lang[1,2], Torbjörn Lundh[3,4], and Max Ortiz-Catalan[1,2,5,6,*]

[1] Center for Bionics and Pain Research, Sweden.

[2] Department of Electrical Engineering, Chalmers University of Technology, Sweden.

[3] Department of Mathematical Sciences, Chalmers University of Technology, Sweden.

[4] Department of Mathematical Sciences, University of Gothenburg, Sweden.

[5] Sahlgrenska Academy, University of Gothenburg, Sweden.

[6] Sahlgrenska University Hospital, Sweden.

*Corresponding author. Address: Department of Electrical Engineering, Chalmers University of Technology, Hörsalsvägen 11, SE – 412 96, Gothenburg, Sweden. Tel.: +46 317725149. Email address: maxo@chalmers.se (M. Ortiz-Catalan).

## Abstract

There is no single prevailing theory of pain that explains its origin, qualities, and alleviation. Although many studies have investigated various molecular targets for pain management, few have attempted to examine the etiology or working mechanisms of pain through mathematical or computational techniques. In this systematic review, we identified mathematical and computational approaches for characterizing pain. The databases queried were *Science Direct* and *PubMed*, yielding 560 articles published prior to January 1$^{st}$, 2020. After screening for inclusion of mathematical or computational models of pain, 31 articles were deemed relevant. Most of the reviewed articles utilized classification algorithms to categorize pain and no-pain conditions. We found the literature heavily focused on the application of existing models or machine learning algorithms to identify the presence or absence of pain, rather than to explore features of pain that may be used for diagnostics and treatment. Although understudied, the development of mathematical models may augment the current understanding of pain by providing directions for testable hypotheses of its underlying mechanisms.

## Introduction

Pain is a subjective experience mediated by a variety of physiological and psychological factors. A personal painful experience may not be easy to communicate and may not be obvious to an observer, however, this lack of apparent objective evidence cannot negate the need for relief. Pain can exist without a physical stimulus, such as with neuropathic pains, and noxious stimuli do not always produce a painful experience, as seen in individuals with congenital insensitivity to pain [39]. This has led to a definition of pain that does not tie it to a physical stimulus. Instead, the International Association for Study of Pain (IASP) describes pain as "an unpleasant sensory and emotional experience associated with actual or potential tissue damage, or described in terms of such damage" [36].

There are many research efforts aiming to understand pain in order to develop pain relief strategies, particularly by using pharmaceutical methods [15]. Animal experiments dealing with pain have suggested possible pain pathways, regulatory systems, and modulatory schemes that may also exist in humans [4,6]. Other attempts to alleviate pain include unconventional techniques such as hypnosis and acupuncture [2,19,40,56]. However, a rigorous scientific understanding of pain and its underlying mechanism have remained out of reach.

As with many other phenomena, pain can be studied at different scales. At the cellular level, hundreds of distinct changes had been identified as potential mediators of neuropathic pain, thus creating an equally large and challenging number of hypotheses yet to be tested [15]. A higher level of study would be that of neural network dynamics, which although encompasses cellular changes as reflected by neural firing behavior, does not require a complete understanding of the cellular changes themselves to utilize the resulting behavior in order to form working hypotheses. To this effect, mathematical and computational models of neural dynamics resulting in pain could be used to elucidate its etiology and serve as guidance for its treatment. Models are descriptions, abstract or material, that reflect or represent, and hence provide access to, selected parts of reality [22]. In order for a model to guide clinical treatment, the model has to give insight into the condition one wants to change. Hence, the focus should be on the understandability power of a model, rather than the usual predictive power as in the current state of pain research. In this work, we examine the efforts made to understand pain via said mathematical and computational models.

Until the second half of the 20$^{th}$ century, two main theories on pain were the *specificity theory* and the *pattern theory*. In 1662, Descartes suggested that pain was a product of neural processing and distinct from nociception [38]. He proposed that noxious stimuli were conveyed to the brain via hollow tubules, and a stimulus of adequate strength would evoke a painful sensation, while a weaker one would evoke a tingling or tickling sensation. Descartes' theory of pain has since been elaborated and its details have culminated under the *specificity theory*, which suggests that stimulation of pain receptors produces nerve impulses that are transmitted to a pain center in the brain via pain-specific pathways. Under this theory, pain is considered an independent sensation, therefore requiring a separate sensory system for its perception—like for vision and hearing [38]. In contrast, *pattern theory* emerged through an effort to quantify sensation; the spatial-temporal pattern of impulses from the peripheral nerves encoded the type of sensation and its intensity. Weak and strong stimuli of the same modality could produce different patterns, thereby producing non-painful or painful experiences. This theory proposed that the central nervous system decoded these impulse patterns, but a sound explanation for this mechanism has not been made [38,48].

In 1965, Melzack and Wall [35] proposed the *gate control theory* of pain, in which spatial-temporal impulse patterns transmitted from peripheral afferent Aβ, Aδ, and C fibers are modulated at the spinal cord, and the modulated signals determine pain modality in the transmission cells in the dorsal horn, from which pain sensation is projected towards the brain. The inclusion of the brain as an important component in pain perception was a major contribution of this theory, as stated by Melzack "never again, after 1965, could anyone try to explain pain exclusively in terms of peripheral factors" [34].

In the *gate control theory* of pain, nociceptive pain is mediated by unmyelinated C fibers. The slightly larger and thinly myelinated Aδ fibers mediate intense, acute pain. The largest nerve fibers—Aβ fibers—respond to touch, pressure, and tension. The *gate control theory* of pain suggests that Aβ fibers also play a role in pain, namely, in pain alleviation. This is exemplified in the short-term reduction of pain by applying light pressure to the painful region, like rubbing a stubbed toe, or applying pressure to the skin after removing a sticky bandage. Sufficient noxious input induces the firing of nociceptors, which can be achieved with appropriate mechanical, thermal, or electrical stimuli [13,33].

No current theory is yet considered to account for all the intricacies of the experience of pain [38]. New ideas continue to emerge including other factors considered important to pain as a multidimensional experience, such as the mature organism model [23] or the dynamic pain connectome [29]. However, little guidance has been provided on how such ideas could be falsified, or experimentally verified.

Our group is particularly interested in phantom limb pain (PLP), a class of neuropathic pain commonly suffered after amputation of an extremity. Several hypotheses on the genesis of PLP were proposed over the last decades that are now challenged by clinical observations [42]. This motivated Ortiz-Catalan to conjecture the *stochastic entanglement hypothesis* for the genesis of PLP, in which pain and sensorimotor circuitry becomes pathologically linked to activate despite the lack of nociceptive input [42]. The somatosensory neural network is intrinsically linked to pain processing because all pain is embodied, *i.e.*, somatosensory processing provides pain of a perceived location. The "entanglement" was hypothesized to initiate by incidental random firing of neurons belonging to the impaired sensorimotor network, unintentionally triggering neurons involved in pain perception. This and other ideas on the etiology and working mechanisms of pain, such as the more general *gate control theory* by Melzack [33] could be potentially studied using mathematical models and computational approaches. The aim of this article is to review and classify published efforts in this direction in order to inform future research.

## Methods

We performed a systematic literature review using the two databases: *Science Direct* and *PubMed*. Article title, key words, and abstracts were searched using the following search condition: (computational biology OR neural network OR mathematical model OR dynamical systems) AND pain AND (perception OR processing OR neuropathic OR chronic OR phantom limb). The inclusion criteria required the articles to contain a mathematical theory or computational approach to characterizing pain. We considered journal articles published prior January 1st, 2020. Conference proceedings, book chapters, editorial letters, and non-English articles were excluded. The screening procedure is presented in detail in Fig. 1.

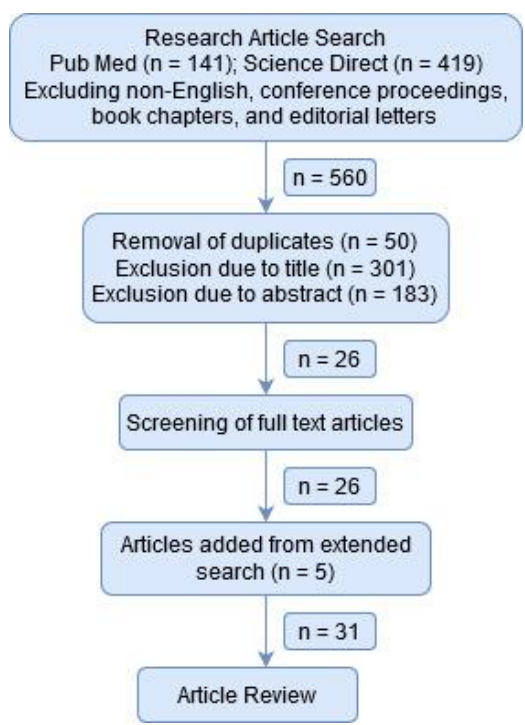

*Figure 1 Schematic view of the methodology used for the systematic review. From the filtered search, the articles reviewed were required to contain a mathematical theory or computational approach to characterizing pain.*

The filtered search yielded 31 unique and relevant articles from 560 initially screened. Relevant articles discussed computational techniques for quantifying pain using clinical data and experiments, and computer simulations to replicate pain processing and perception. The peer-reviewed research articles were studied to identify the current mathematical and computational approaches to studying pain, and theories regarding the generation, qualities, and alleviation of pain. A breakdown of article types is given in Fig. 2.

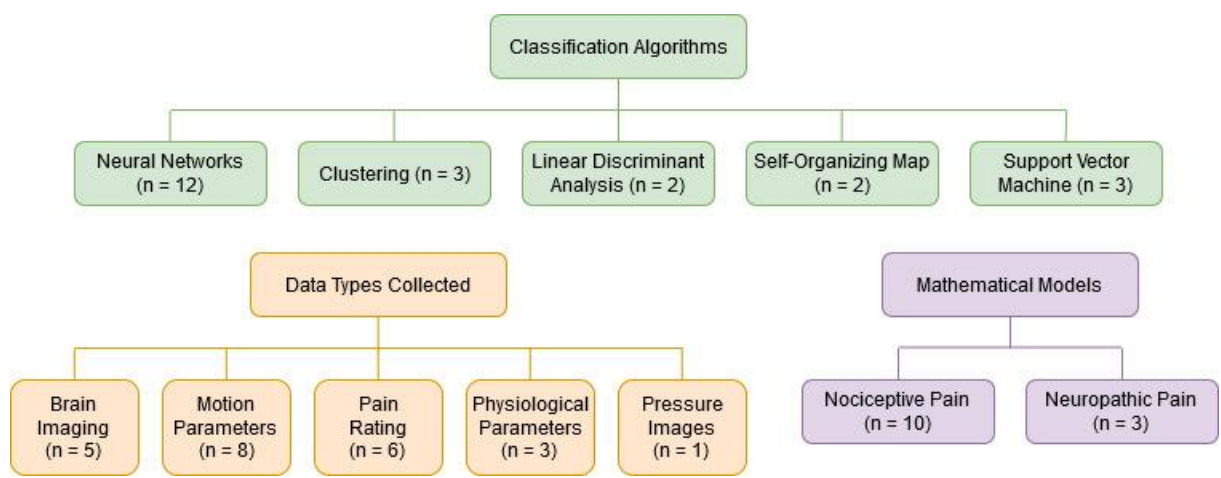

*Figure 2 Articles sorted according to their classification algorithm, data collection method, or proposal of a mathematical model. Articles could belong to more than one category and subcategory.*

## Results

### *Mathematical Models of Pain*

Table 1 lists, classifies (according to the model taxonomy in [22]), and describes all articles found presenting mathematical models of pain. Ten articles developed models for nociceptive pain and three for neuropathic pain.

We found that the first quantitative analysis of nociceptive conduction was performed in 1981. Minamitani and Hagita proposed a mathematical model that generated a numerical description of nociceptive pain and touch sensations [37]. Based on findings in physiological and anatomical literature, including the gate control theory schematic from Melzack and Wall [35], the model simulated a one-directional ascending and descending pathway for pain sensation. Peripheral receptors, afferent Aβ, Aδ, and C fibers, and receptive neurons of the spinal cord, brain stem, thalamus, and the cerebral cortex were considered. To reduce complexity, interactions from lateral inhibition and facilitation were not included in the model. Even so, the model ended up with over 70 parameters. Whereas adaptation and conduction velocity of the fibers were considered, the fibers in each neural unit, consisting of the afferent fiber types, were prescribed with constant conduction velocity and firing threshold. The simulation was conducted with a single square-wave pulse and a periodic repetitive pulse applied to peripheral receptors. The activities of the neurons in the periphery and the upper brain were represented by Wilson-Cowan's nonlinear differential equation. See the coupled pair of ordinary differential equations (ODE) (11) and (12) in [53], which considers continuous neuronal activity, and the distribution of peripheral receptors were described as Gaussian. This system was able to generate hysteresis and limit cycles; see also [52] for further details. The firing characteristics of the neurons were compared to physiological findings, where the results of the simulation and literature coincided satisfactorily. The modality of graded touch sensation, "fast stinging pain" mediated by small unmyelinated Aδ fibers, and "slow burning pain" mediated by unmyelinated C fibers were successfully simulated, despite the simplification of the model. This work suggested that this proposed neural network was useful to characterize different sensory modalities in pain.

In 1989, Britton and Skevington translated the *gate control theory* of pain [35] into a mathematical model simulating acute pain for a single transmission unit, consisting of one Aβ fiber, one C fiber, one inhibitory neuron, and one excitatory neuron [8]. Fig. 3 shows the gate control theory adapted schematically. They proposed a system of partial differential equations that exhibited all characteristics of acute pain described in the theory using the above mentioned Wilson-Cowan model of activity in a synaptically coupled neuronal network [53]. Furthermore, with some variation of parameters, they suggested that their model could demonstrate temporal pain qualities like throbbing and pulsing, and possibly expand the model's validity to include neuropathy. This is quite remarkable given that the model is relatively minimalistic. See Equations (9) in [8] that consist of three coupled ODEs and only three unknown variables. This was utilized by the authors to prove analytic results about the uniqueness of a steady state solution. In 2004, this model was repeated and expanded, using more modern computational hardware and software, for the assumption that neighboring transmission units behave similarly [44]. The acute pain model was extended by increasing the number of transmission units communicating to the midbrain. Neighboring transmission units were found to behave similarly, but the transmission unit potential decreased when the number of transmission units increased sufficiently. This suggested a saturation point in which transmission units may fail to fire despite neural fiber activation [44].

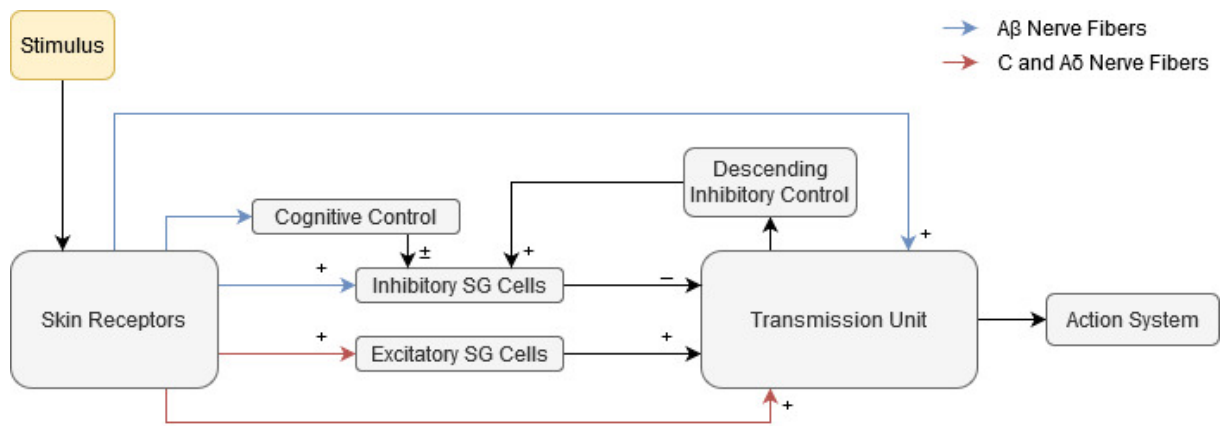

*Figure 3 Melzack and Wall's gate control theory of pain shown schematically. Compare with the original Figure 4 in [54] and the variants in Figure 1 in [8] and Figure 1 in [41]. Plus signs (+) denote excitation, and minus signs (-) denote inhibition. Cognitive control can be excitatory or inhibitory. SG = substantia gelatinosa cells in the dorsal horn of the spinal cord.*

In 2012, Rho and Prescott used relatively sophisticated tools and insights in non-linear dynamics to develop a computational model to simulate the onset of neuronal hyperexcitability from a normal spiking pattern [45]. Nerve injuries may cause various molecular changes, and any one change may singly cause neuropathy. The model demonstrated that the accumulation of small parameter changes was enough for modifying the normal spiking pattern to a repetitive one, enabling membrane potential oscillations, and bursting. These changes suggested that all three pathological events are related. This study contrasts with other studies on neuropathic pain in that instead of identifying changes that result from pain, this model examined the combination of molecular changes that result in neuropathic excitability.

In 2014, Boström *et al.* developed a computational model of PLP, suggesting that phantom pain, maladaptive reorganization, and persistent representation are all symptoms of the same underlying mechanism that results in ectopic spontaneous activity of deafferented nociceptive channels [7]. They considered the somatosensory cortex to be dynamically self-organized, where spontaneous activity in the sensory system exists and is abnormally increased in regions affected by deafferentation. Following the *gate control theory*'s discussion on pain sensitization, they presume that a long-term decrease in input strength results in a similar change to the gating threshold. Their results suggest that amputation causes a threshold decrease, thereby allowing greater spontaneous activity in the sensory system. The assumptions required for this computational model consider nociception, but they excluded an explanation about a possible mechanism for pain perception or instantiation of a painful experience. Their model was based on PLP being the result of increased spontaneous nociceptive firing in the severed nerves, and thus initiated and maintained by the peripheral nervous system, which contradicts current views on PLP being maintained by central changes [20,21].

*Pain classification algorithms*

A considerable number of studies on pain involved pain questionnaires, experimental setups, and brain imaging. A small subset of these have used classification algorithms for identifying characteristics of pain from collected data. Table 2 lists, classifies, and describes the 18 articles found in this systematic review that relied on classification algorithms to differentiate between the presence or absence of pain.

Twelve of the 18 used neural networks, and 6 of these were applied to identify chronic low back pain using parameters acquired from surface electromyography (sEMG), kinematic data, images, or video. Classification accuracies measured 80% or higher, indicating neural networks could be complementary to physiotherapist assessments [11,18,24,27,32,41]. Two additional articles utilized self-organizing maps to categorize patients with and without chronic low back pain from activity levels and pain questionnaires [30,31].

In another effort to autonomously classify data, Atlas *et al.* used cluster analysis with fMRI, demonstrating some brain regions responded to the intensity of a thermal stimulus but did not predict pain [1]. Other brain regions did not respond to noxious input but did predict pain. Balaban *et al.* used fuzzy C-means cluster analysis to identify three response phenotypes to two time-varying oral capsaicin administration paradigms in order to examine individual pain differences [3]. This work proposed classification of human pain responses by temporal pattern, rather than by threshold or magnitude of response.

## Discussion

A vast majority of research on pain does not attempt to understand pain in all its varieties, but rather many efforts are made toward understanding factors that contribute to specific pain conditions. Often these efforts require experimental laboratory work or clinical studies which can have long reporting times and are unique to the molecules explored or subject data collected. Mathematical models are advantageous in that they do not require data acquisition and can be designed to examine specific or general phenomena. Despite these benefits, the publications we found on mathematical modelling of pain averaged 16.8 citations (±13.5) by March 31$^{st}$, 2020, which is an indication that mathematical modelling has not been a tool widely employed nor accepted in the study of pain. This is notwithstanding its proven utility in helping to understand complex biological processes, such as in exploring the neural dynamics of vision and in predicting the spread of infectious diseases [46,54].

A possible explanation for the lack of mathematical models on pain is the lack of appropriate data to test them. There are many molecular candidates for targeted pain relief [15], but their interactions and dynamics are difficult to inject into a mathematical theory for pain without creating a model unique to a single set of molecules and their conditions. There is an ever-growing number of parameters to consider, and the complexity of developing a robust model that does not oversimplify the pain experience is a mounting barrier; see for example [22]. In Minamitani and Hagita's model [37], even without considering lateral inhibition and facilitation, their model utilizes over 70 parameters. The sheer number of parameters complicates model construction such that it is a deterrent to replications and expansions. Furthermore, the psychological factors that affect the pain experience are difficult to quantify, and their influence on the pain experience is unclear. At this point, it is unclear which biomarkers are present in all instances of pain, hence the difficulty in creating an all-inclusive model.

We found that the few mathematical models that tackled the root of pain generation often relied on simplified models of neural mechanism and missed the mark of mimicking the variety of pain types. In Britton and Skevington's 1989 mathematical model of the *gate control theory*, acute pain was successfully simulated and further analysis suggested an explanation for temporal qualities of pain, a phenomenon previously unexplained by this theory [8]. However elegantly constructed from a set of three coupled ODEs, the parameters and functions were only generally described and essentially not based on experimental measurements. The possibility of modifying the model to also simulate chronic pain via the inclusion of a plasticity scheme was surprisingly never realized in any future work [25,44], despite its tangibility. Efforts towards a general understanding of pain and its origin diminished in more recent decades, in favor of models for specific pain conditions.

Recent models have examined specific aspects of pain, such as daily variance of intensity, nociceptor response to chemical or thermal stimuli, and PLP [7,14,16,17,55]. The limitation of these models lies in the lack of generalizability. The computational model of PLP developed by Boström *et al.* [7] is one that claims that spontaneous activity in the somatosensory cortex is abnormally increased in regions affected by deafferentation, in which PLP arises. In line with mechanisms of pain sensitization from the *gate control theory*, they assume that amputation puts no constraint on nociceptive input due to the lack of sensory input, leading to a decrease in gating threshold and thereby an increase in spontaneous activity. Self-organization of the somatosensory cortex would maladapt to this sustained absence of sensory information and result in PLP [7]. Although nociception is considered in this model, no pain mechanism is discussed that supports maintenance of firing at the severed nerves. Despite the complexity of this model, the results were expected given that their model was constructed using rather strong assumptions on the underlying PLP process and well-known but complex phenomenologically model parts, such as Kohonen maps (also known as self-organizing maps, SOM).

Articles discussing pain classification algorithms averaged 21.7 citations (SD = 25.6). Most of this literature focused on categorizing behavioral and brain imaging data to identify the presence of pain. The primary focus of these articles was on developing more effective methods of identifying characteristics and patterns of pain in order to give more accurate predictions.

The use of artificial neural networks was successful for identifying patients with chronic low back pain using motion parameters and sEMG, suggesting its diagnostic power. However, none of these articles attempted to differentiate no pain, acute pain, and chronic pain due to limitations set on the scope of the experiments. The examination of pain features was not conducted to explore pain characteristics that are difficult to observe in a clinical setting. In studies comparing clinician and algorithmic predictive successes, there were missed opportunities to scrutinize examples in which the clinician and algorithm disagreed. This error analysis could have been used to improve the algorithms and to augment the clinician's knowledge in diagnosing pain conditions. Lastly, the reliance on collecting suitable data types in sufficient amounts is detrimental when relying on these classification methods. Despite these challenges, computational models are invaluable tools for examining the full spectrum of conditions because combinations of parameters can be studied faster than with clinical studies, which cannot test exhaustively.

Researchers are encouraged to provide experimental procedures to falsify their proposed hypothesis, such as those illustrated by Devor on molecular mechanisms [15] and Ortiz-Catalan on PLP [42].

## Acknowledgements

This research was supported by the Promobilia Foundation, the IngaBritt and Arne Lundbergs Foundation, Vinnova, the European Pain Federation–Grünenthal–Research Grant (E-G-G) and the European Commission.

Conflict of Interest: The first and second authors report no conflict of interest. The last author served as a consultant for Integrum AB.

## References

[1] Atlas LY, Lindquist MA, Bolger N, Wager TD. Brain mediators of the effects of noxious heat on pain. Pain 2014;155:1632–1648. doi:10.1016/j.pain.2014.05.015.

[2] Bai L, Qin W, Tian J, Dai J, Yang W. Detection of dynamic brain networks modulated by acupuncture using a graph theory model. Prog Nat Sci 2009;19:827–835. doi:10.1016/j.pnsc.2008.09.009.

[3] Balaban CD, McBurney DH, Affeltranger MA. Three distinct categories of time course of pain produced by oral capsaicin. J Pain 2005;6:315–322. doi:10.1016/j.jpain.2005.01.346.

[4] Becerra L, Bishop J, Barmettler G, Kainz V, Burstein R, Borsook D. Brain network alterations in the inflammatory soup animal model of migraine. Brain Res 2017;1660:36–46. doi:10.1016/j.brainres.2017.02.001.

[5] Behrman M, Linder R, Assadi AH, Stacey BR, Backonja M-M. Classification of patients with pain based on neuropathic pain symptoms: Comparison of an artificial neural network against an established scoring system. Eur J Pain 2007;11:370–376. doi:10.1016/j.ejpain.2006.03.001.

[6] Borsook D, Becerra L. CNS animal fMRI in pain and analgesia. Neurosci Biobehav Rev 2011;35:1125–1143. doi:10.1016/j.neubiorev.2010.11.005.

[7] Boström KJ, De Lussanet MHE, Weiss T, Puta C, Wagner H. A computational model unifies apparently contradictory findings concerning phantom pain. Sci Rep 2014;4:5298. doi:10.1038/srep05298.

[8]   Britton NF, Skevington SM. A mathematical model of the gate control theory of pain. J Theor Biol 1989;137:91–105. doi:10.1016/S0022-5193(89)80151-1.

[9]   Brodersen KH, Wiech K, Lomakina EI, Lin C shu, Buhmann JM, Bingel U, Ploner M, Stephan KE, Tracey I. Decoding the perception of pain from fMRI using multivariate pattern analysis. Neuroimage 2012;63:1162–1170. doi:10.1016/j.neuroimage.2012.08.035.

[10]  Cannistraci CV, Ravasi T, Montevecchi FM, Ideker T, Alessio M. Nonlinear dimension reduction and clustering by Minimum Curvilinearity unfold neuropathic pain and tissue embryological classes. Bioinformatics 2011;27:i531–i539. doi:10.1093/bioinformatics/btq376.

[11]  Caza-Szoka M, Massicotte D, Nougarou F, Descarreaux M. Surrogate analysis of fractal dimensions from SEMG sensor array as a predictor of chronic low back pain. Proceedings of the Annual International Conference of the IEEE Engineering in Medicine and Biology Society, EMBS. IEEE, 2016, Vol. 2016- Octob. pp. 6409–6412. doi:10.1109/EMBC.2016.7592195.

[12]  Cecchi GA, Huang L, Hashmi JA, Baliki M, Centeno M V., Rish I, Apkarian AV. Predictive Dynamics of Human Pain Perception. PLoS Comput Biol 2012;8:e1002719. doi:10.1371/journal.pcbi.1002719.

[13]  Cervero F, Merskey H. What is a noxious stimulus? Pain Forum 1996;5:157–161. doi:10.1016/S1082-3174(96)80020-1.

[14]  Crodelle J, Piltz SH, Hagenauer MH, Booth V. Modeling the daily rhythm of human pain processing in the dorsal horn. PLOS Comput Biol 2019;15:e1007106. doi:10.1371/journal.pcbi.1007106.

[15]  Devor M. Neuropathic pain: what do we do with all these theories? Acta Anaesthesiol Scand 2001;45:1121–1127. doi:10.1034/j.1399-6576.2001.450912.x.

[16]  Dick OE. Mechanisms of dynamical complexity changes in patterns of sensory neurons under antinociceptive effect emergence. Neurocomputing 2020;378:120–128. doi:10.1016/j.neucom.2019.10.004.

[17]  Dick OE, Krylov B V., Nozdrachev AD. Possible mechanism of bursting suppression in nociceptive neurons. Dokl Biochem Biophys 2017;473:137–140. doi:10.1134/S1607672917020120.

[18]  Dickey JP, Pierrynowski MR, Bednar DA, Yang SX. Relationship between pain and vertebral motion in chronic low-back pain subjects. Clin Biomech 2002;17:345–352. doi:10.1016/S0268-0033(02)00032-3.

[19]  Faymonville M-E, Roediger L, Del Fiore G, Delgueldre C, Phillips C, Lamy M, Luxen A, Maquet P, Laureys S. Increased cerebral functional connectivity underlying the antinociceptive effects of hypnosis. Cogn Brain Res 2003;17:255–262. doi:10.1016/S0926-6410(03)00113-7.

[20]  Flor H, Andoh J. Origin of phantom limb pain: A dynamic network perspective. e-Neuroforum 2017;23:111–116. doi:10.1515/nf-2017-A018.

[21]  Flor H, Nikolajsen L, Staehelin Jensen T. Phantom limb pain: a case of maladaptive CNS plasticity? Nat Rev Neurosci 2006;7:873–881. doi:10.1038/nrn1991.

[22] Gerlee P, Lundh T. Scientific Models: Red Atoms, White Lies and Black Boxes in a Yellow Book. Cham: Springer International Publishing, 2016 p. doi:10.1007/978-3-319-27081-4.

[23] Gifford L. Pain, the Tissues and the Nervous System: A conceptual model. Physiotherapy 1998;84:27–36. doi:10.1016/S0031-9406(05)65900-7.

[24] Gioftsos G, Grieve DW. The use of artificial neural networks to identify patients with chronic low-back pain conditions from patterns of sit-to-stand manoeuvres. Clin Biomech 1996;11:275–280. doi:10.1016/0268-0033(96)00013-7.

[25] Haeri M, Asemani D, Gharibzadeh S. Modeling of pain using artificial neural networks. J Theor Biol 2003;220:277–284. doi:10.1006/jtbi.2003.3130.

[26] Henssen DJHA, Witkam RL, Dao JCML, Comes DJ, Van Cappellen van Walsum A-M, Kozicz T, van Dongen R, Vissers K, Bartels RHMA, de Jong G, Kurt E. Systematic Review and Neural Network Analysis to Define Predictive Variables in Implantable Motor Cortex Stimulation to Treat Chronic Intractable Pain. J Pain 2019;20:1015–1026. doi:10.1016/j.jpain.2019.02.004.

[27] Hu B, Kim C, Ning X, Xu X. Using a deep learning network to recognise low back pain in static standing. Ergonomics 2018;61:1374–1381. doi:10.1080/00140139.2018.1481230.

[28] Keijsers NLW, Stolwijk NM, Louwerens JWK, Duysens J. Classification of forefoot pain based on plantar pressure measurements. Clin Biomech 2013;28:350–356. doi:10.1016/j.clinbiomech.2013.01.012.

[29] Kucyi A, Davis KD. The dynamic pain connectome. Trends Neurosci 2015;38:86–95. doi:10.1016/j.tins.2014.11.006.

[30] Liszka-Hackzell JJ, Martin DP. Analysis of nighttime activity and daytime pain in patients with chronic back pain using a self-organizing map neural network. J Clin Monit Comput 2005;19:411–414. doi:10.1007/s10877-005-0392-8.

[31] Liszka-Hackzell JJ, Martin DP. Categorization and analysis of pain and activity in patients with low back pain using a neural network technique. J Med Syst 2002;26:337–347.

[32] Magnusson ML, Bishop JB, Hasselquist L, Spratt KF, Szpalski M, Pope MH. Range of motion and motion patterns in patients with low back pain before and after rehabilitation. Spine (Phila Pa 1976) 1998;23:2631–2639. doi:10.1097/00007632-199812010-00019.

[33] Melzack R. Gate control theory. Pain Forum 1996;5:128–138. doi:10.1016/S1082-3174(96)80050-X.

[34] Melzack R. Pain - an overview. Acta Anaesthesiol Scand 1999;43:880–884. doi:10.1034/j.1399-6576.1999.430903.x.

[35] Melzack R, Wall PD. Pain Mechanisms: A New Theory. Science (80- ) 1965;150:971–978. doi:10.1126/science.150.3699.971.

[36] Merskey H, Albe Fessard D, Bonica JJ, Carmon A, Dubner R, Kerr FWL, Lindblom U, Mumford JM, Nathan PW, Noordenbos W, Pagni CA, Renaer MJ, Sternbach RA SS. Pain terms: a list with definitions and notes on usage. Recommended by the IASP

Subcommittee on Taxonomy. Pain 1979;6:249–252.

[37] Minamitani H, Hagita N. A Neural Network Model of Pain Mechanisms: Computer Simulation of the Central Neural Activities Essential for the Pain and Touch Sensations. IEEE Trans Syst Man Cybern 1981;11:481–493. doi:10.1109/TSMC.1981.4308725.

[38] Moayedi M, Davis KD. Theories of pain: from specificity to gate control. J Neurophysiol 2013;109:5–12. doi:10.1152/jn.00457.2012.

[39] Nagasako EM, Oaklander AL, Dworkin RH. Congenital insensitivity to pain: an update. Pain 2003;101:213–219. doi:10.1016/S0304-3959(02)00482-7.

[40] Napadow V, Dhond RP, Kim J, LaCount L, Vangel M, Harris RE, Kettner N, Park K. Brain encoding of acupuncture sensation — Coupling on-line rating with fMRI. Neuroimage 2009;47:1055–1065. doi:10.1016/j.neuroimage.2009.05.079.

[41] Oliver CW, Atsma WJ. Artificial intelligence analysis of paraspinal power spectra. Clin Biomech 1996;11:422–424. doi:10.1016/0268-0033(96)00016-2.

[42] Ortiz-Catalan M. The Stochastic Entanglement and Phantom Motor Execution Hypotheses: A Theoretical Framework for the Origin and Treatment of Phantom Limb Pain. Front Neurol 2018;9:1–16. doi:10.3389/fneur.2018.00748.

[43] Ozkan O, Yildiz M, Arslan E, Yildiz S, Bilgin S, Akkus S, Koyuncuoglu HR, Koklukaya E. A Study on the Effects of Sympathetic Skin Response Parameters in Diagnosis of Fibromyalgia Using Artificial Neural Networks. J Med Syst 2016;40:54. doi:10.1007/s10916-015-0406-0.

[44] Prince K, Campbell J, Picton P, Turner S. A computational model of acute pain. Int J Simul Syst Sci Technol 2005;6:1–10.

[45] Rho YA, Prescott SA. Identification of molecular pathologies sufficient to cause neuropathic excitability in primary somatosensory afferents using dynamical systems theory. PLoS Comput Biol 2012;8:e1002524. doi:10.1371/journal.pcbi.1002524.

[46] Said CP, Heeger DJ. A Model of Binocular Rivalry and Cross-orientation Suppression. PLoS Comput Biol 2013;9:e1002991. doi:10.1371/journal.pcbi.1002991.

[47] Santana AN, Cifre I, de Santana CN, Montoya P. Using Deep Learning and Resting-State fMRI to Classify Chronic Pain Conditions. Front Neurosci 2019;13:1–13. doi:10.3389/fnins.2019.01313.

[48] Sinclair DC. Cutaneous sensation and the doctrine of specific energy. Brain 1955;78:584–614. doi:10.1093/brain/78.4.584.

[49] Spitzer M, Böhler P, Weisbrod M, Kischka U. A neural network model of phantom limbs. Biol Cybern 1995;72:197–206. doi:10.1007/BF00201484.

[50] Tigerholm J, Petersson ME, Obreja O, Lampert A, Carr R, Schmelz M, Fransén E. Modeling activity-dependent changes of axonal spike conduction in primary afferent C-nociceptors. J Neurophysiol 2014;111:1721–1735. doi:10.1152/jn.00777.2012.

[51] Vuckovic A, Gallardo VJF, Jarjees M, Fraser M, Purcell M. Prediction of central neuropathic pain in spinal cord injury based on EEG classifier. Clin Neurophysiol 2018;129:1605–1617. doi:10.1016/j.clinph.2018.04.750.

[52]   Wilson HR, Cowan JD. A mathematical theory of the functional dynamics of cortical and thalamic nervous tissue. Kybernetik 1973;13:55–80. doi:10.1007/BF00288786.

[53]   Wilson HR, Cowan JD. Excitatory and Inhibitory Interactions in Localized Populations of Model Neurons. Biophys J 1972;12:1–24. doi:10.1016/S0006-3495(72)86068-5.

[54]   Wu JT, Cowling BJ. The use of mathematical models to inform influenza pandemic preparedness and response. Exp Biol Med 2011;236:955–961. doi:10.1258/ebm.2010.010271.

[55]   Xu F, Wen T, Seffen K, Lu T. Modeling of skin thermal pain: A preliminary study. Appl Math Comput 2008;205:37–46. doi:10.1016/j.amc.2008.05.045.

[56]   Zhao Z-Q. Neural mechanism underlying acupuncture analgesia. Prog Neurobiol 2008;85:355–375. doi:10.1016/j.pneurobio.2008.05.004.

1 # Tables

2 *Table 1 Literature on Mathematical Models of Pain*

| Publication | Model types [53] | Summary |
|---|---|---|
| Minamitani and Hagita (1981) [37] | Analogous/ symbolic | The neural network model simulated the conduction mechanism of pain and touch sensations. Although only one directional ascending and descending pathway for pain sensation was represented, and no interaction from inhibition or facilitation was considered, the modalities of graded touch sensation and two different pain modalities were observed. |
| Britton and Skevington (1989) [8] | Analogous/ symbolic | Melzack's gate control theory of pain was translated into a mathematical model simulating acute pain for a single transmission unit. The partial differential equations were based on the Wilson-Cowan model for synaptically coupled neuronal networks. |
| Spitzer *et al.* (1995) [49] | Analogous/ phenomenological | A self-organization feature map using Kohonen network was used to simulate the effects of amputation. The Kohonen network was trained on input patterns and subsequently deprived parts of the input patterns in order to simulate partial deafferentation. This led to reorganization driven by input noise, which represented noise generated by erratic firing of lacerated dorsal root ganglion sensory neurons. |
| Haeri *et al.* (2003) [25] | Analogous/ phenomenological | An artificial neural network to model the steady state behavior of pain mechanisms was developed using input patterns from small and large nerve fibers. For stimulation states corresponding to acute pain, a collection of basic patterns was used as features for the model. Given a novel pain stimulus, the prediction of pain was possible. |
| Xu *et al.* (2008) [55] | Symbolic | Considering the biophysical and neural mechanisms of pain sensation, a mathematical model for quantifying skin thermal pain that included transduction, transmission, and perception was proposed. This model proposed that the intensity of thermal pain was related to the character of the noxious stimulus. |
| Cecchi *et al.* (2012) [12] | Analogous/ symbolic | Thermal pain perception was modelled as a dynamical system to be compared to reported pain ratings from intensity-varying thermal stimuli. Using a sparse regression method, pain ratings were predicted according to fMRI data and reported pain ratings. |
| Rho and Prescott (2012) [45] | Conceptual/ symbolic | A computational model was developed to simulate the onset of neuronal hyperexcitability from a normal spiking pattern. Parameters changes were sufficient to alter the normal spiking pattern to a repetitive one, enabling membrane potential oscillations, and bursting, suggesting that the three pathologies are related. |
| Boström *et al.* (2014) [7] | Conceptual/ phenomenological | A computational model of phantom limb pain was developed based on the increase of spontaneous nociceptive firing. They proposed that the same underlying mechanism that results in ectopic spontaneous activity of deafferented nociceptive channels was responsible for phantom pain, maladaptive reorganization, and persistent representation. |
| Prince *et al.* (2014) [44] | Analogous/ symbolic | Britton and Skevington's acute pain model was replicated and expanded to verify the assumption that neighboring transmission units behave similarly. With sufficient increase in the number of transmission units input to the midbrain, transmission unit potential decreased, suggesting a saturation point in which transmission units may fail to fire despite neural fiber activation. |
| Tigerholm *et al.* (2014) [50] | Analogous/ symbolic | Axonal conduction velocity by activity differs between patients with neuropathic pain and those without, suggesting that this property may play a role in the development of neuropathies. A mathematical model of human cutaneous C-fibers was developed to investigate the activity-dependent changes of axonal spike conduction. |
| Dick *et al.* (2017) [17] | Analogous/ symbolic | By implementing a mathematical model of rat nociceptive neuronal membrane, a mechanism of ectopic bursting suppression in dorsal root ganglia neurons with comenic acid was proposed. The administration of comenic acid to the model reduced rhythmic discharge frequency due to a decrease in the effective charge transferring via sodium gate activation dynamics. |
| Crodelle *et al.* (2019) [14] | Statistical/ symbolic | A mathematical model of the dorsal horn neural circuit relying on firing rates and model parameters from experimental literature was developed to describe daily modulation of pain sensitivity. The inversion of daily rhythmicity of pain in neuropathic patients was proposed to be the result of dorsal horn circuitry dysregulation. |
| Dick (2020) [16] | Analogous/ symbolic | Bifurcation analysis was used to determine the relationship between the nociceptive neuron model and the antinociceptive effect that occurs during neuropathic pain suppression. The molecular mechanism of the bursting suppression was associated with the modification of the activation gating system of Nav1.8 channels by comenic acid, suggesting a possible molecular treatment for neuropathic pain. |

*Table 2 Literature on Pain Classification Algorithms*

| Publication | Classification Algorithm | Summary |
|---|---|---|
| Gioftsos and Grieve (1996) [24] | ANN | Categorized chronic low back pain patients, fake low back pain patients, and healthy controls based on sit-to-stand maneuvers using an ANN and physiotherapist assessment. The ANN better detected abnormal movement patterns but was not necessarily better at diagnosing. |
| Oliver and Atsma (1996) [41] | ANN | Categorized chronic low back pain patients and healthy controls using sEMG power spectra data collected from contraction tasks using an ANN. |
| Magnusson *et al.* (1998) [32] | ANN | Identified chronic low back pain characteristics from trunk motion data in patients undergoing chronic low back pain rehabilitation. |
| Dickey *et al.* (2002) [18] | ANN, LDA | Chronic low back pain motion, intravertebral deformation, and pain were assessed with LDA and an ANN. The ANN showed a strong relationship between observed and predicted pain due to the nonlinear relationship between vertebral motion parameters and pain. |
| Liszka and Martin (2002) [30] | SOM | Categorized pain and activity levels from patients with acute back pain and chronic back pain. |
| Liszka and Martin (2005) [31] | SOM | Investigated the relationship between daytime chronic back pain levels and sleep activity using a SOM neural network. Results showed that daytime pain levels and sleep activity were not correlated, however, daytime pain variance was correlated with sleep activity levels and patterns. |
| Balaban *et al.* (2005) [3] | Cluster analysis | Identified 3 response phenotypes (level detection, change detection, and cumulative irritation) to 2 time-varying capsaicin administration paradigms and proposed a method of classifying human pain responses by temporal pattern, rather than by threshold or magnitude of response. |
| Behrman *et al.* (2007) [5] | ANN | Compared an ANN to traditional scoring systems for differentiating neuropathic pain and non-neuropathic pain patients using responses from a neuropathic pain questionnaire. |
| Cannistraci *et al.* (2010) [10] | Cluster analysis | Proposed *Minimum Curvilinearity* for dimension reduction and clustering for the classification of 2D electrophoresis gel images derived from proteomic cerebrospinal fluid profiles of peripheral neuropathic patients and ALS patients unaffected by neuropathic pain. This method improved classification accuracy of small nonlinear datasets. |
| Brodersen *et al.* (2012) [9] | SVM | Investigated the predictive ability of fMRI data for decoding painful stimuli using multivariate analysis on different spatial scales (single voxels, individual anatomical regions, combinations of regions, or whole-brain activity). |
| Keijsers *et al.* [28] | ANN | Identified differences in plantar pressure patterns in people with and without forefoot pain. |

| Atlas *et al.* (2014) [1] | Cluster analysis | Identified brain mediators of pain induced by thermal stimuli using multi-level mediation analysis. Cluster analysis showed that the mediators belonged to several distinct functional networks with complementary roles in pain genesis. The identified networks did not necessarily respond to noxious input and predict pain, indicating various brain regions contribute to the pain. |
|---|---|---|
| Ozkan *et al.* (2015) [43] | ANN | Demonstrated that the inclusion of sympathetic skin response parameters as a feature in a fibromyalgia ANN model increases classification accuracy. |
| Caza *et al.* (2016) [11] | ANN | Categorized chronic low back pain patients and healthy controls from sEMG data collected from a muscle endurance task. A surrogate analysis of the data scored each channel on the sEMG sensory array based on the fractal dimension, showing nonlinearity. The most nonlinear values were used as signal characteristics for the ANN model. |
| Hu *et al.* (2018) [27] | ANN | Identified low back pain patients using static balance control performance data during static standing tasks. |
| Vuckovic *et al.* (2018) [51] | ANN, LDA, SVM | Classified spinal cord injured patients at risk of developing neuropathic pain by comparing them to patients who had already developed pain and healthy controls. |
| Henssen *et al.* (2019) [26] | ANN | Identified predictive variables that influence the outcome of implantable motor cortex stimulation for intractable pain. |
| Santana *et al.* (2019) [47] | ANN, SVM | Resting-state fMRI data from chronic pain patients and healthy controls were collected to assess the accuracy of different machine learning models for classification of chronic pain. |

5  ANN = Artificial neural network

6  LDA = Linear discriminant analysis

7  sEMG = Surface electromyography

8  SOM = Self-organizing map

9  SVM = Support vector machine